\documentstyle[12pt,aps,prd,preprint]{revtex}
\begin{document}
\title{Universal Entropy Bound for Rotating Systems}
\author{Shahar Hod}
\address{The Racah Institute for Physics, The
Hebrew University, Jerusalem 91904, Israel}
\date{\today}
\maketitle

\begin{abstract}

We conjecture a {\it universal upper bound} to the entropy of a {\it rotating}
system. The entropy bound follows from application of the generalized
second law of thermodynamics to an idealized gedanken experiment in which an 
entropy-bearing rotating system falls into a black
hole. This bound is {\it stronger} than the Bekenstein entropy
bound for non-rotating systems. 
\end{abstract}
\bigskip

One of the most intriguing features of both the classical and quantum
theory of black-holes is the striking 
analogy between the laws of black-hole physics and the
universal laws of 
thermodynamics \cite{Chris,ChrisRuf,Haw1,Haw2,Car,BarCarHaw}. In
particular, Hawking's (classical) theorem \cite{Haw1}, ``The
surface area of a black hole never decreases,'' is 
a property reminiscent of the entropy of a closed system. This
striking analogy had led Bekenstein \cite{Beken1,Beken2,Beken3} to
conjecture that the area of a black hole (in suitable units) may be
regarded as the black-hole entropy -- entropy in the sense of
information about the black-hole interior inaccessible to observers
outside the black hole. This conjecture is logically related to a
second conjecture, known as {\it the generalized second law of 
thermodynamics} (GSL): ``{\it The sum of the black-hole 
entropy} (now known to be $1 \over 4$ of the horizon's 
surface area) {\it and the common (ordinary) entropy in the black-hole exterior never decreases}''. 

The general belief in the validity of the ordinary second law of
thermodynamics rests
mainly on the repeated failure over the years of attempts 
to violate it. There currently exists no general proof of 
the law based on the known microscopic laws 
of physics.
In the analog case of the GSL considerably less is known
since the fundamental microscopic laws of physics, namely, the
laws of quantum gravity are not yet known. Hence, one is forced to
consider gedanken experiments in order to test the validity of the
GSL. Such experiments are important since the validity of the GSL
underlies the relationship between black-hole physics and
thermodynamics. If the GSL is valid, then it is very plausible that
the laws of black-hole physics are simply the ordinary laws of
thermodynamics applied to a self-gravitating quantum system. This
conclusion, if true, would provide a striking demonstration 
of the {\it unity} of physics. 
Thus, it is of considerable interest to test the validity of the 
GSL in various gedanken experiments.

In a {\it classical} context, a basic physical mechanism is known 
by which a violation of the GSL can be achieved:
Consider a box filled with matter of proper energy $E$ and 
entropy $S$ which is dropped into a black hole. The energy delivered
to the black hole can be arbitrarily {\it red-shifted} by letting the
assimilation point approach the black-hole horizon. As shown by
Bekenstein \cite{Beken3,Beken4}, if the box is deposited with no radial momentum a proper
distance $R$ above the horizon, and then allowed to fall in such that

\begin{equation}\label{Eq1}
R < \hbar S/2 \pi E\  ,
\end{equation}
then the black-hole area increase (or equivalently, the increase in 
black-hole entropy) is not large enough to compensate for the decrease
of $S$ in common (ordinary) entropy. 
Bekenstein has proposed a
resolution of this apparent violation of the GSL which is based on
the {\it quantum} nature of the matter dropped into the black
hole. He has proposed the existence of a universal upper bound on the
entropy $S$ of any system of total energy $E$ and maximal radius $R$ \cite{Beken5}:

\begin{equation}\label{Eq2}
S \leq 2\pi RE/\hbar\  .
\end{equation}
It has been argued \cite{Beken5,Beken6,Beken7}, and disputed
\cite{UnWal,PelWal} that this restriction is {\it necessary} for enforcement of the GSL; the
box's entropy disappears but an increase in black-hole entropy occurs
which ensures that the GSL is respected provided $S$ is bounded 
as in Eq. (\ref{Eq2}). Other derivations of the universal 
bound Eq. (\ref{Eq2}) which are based on black-hole physics have been
given by Zaslavskii\cite{Zas1,Zas2,Zas3} and by Li and Liu
\cite{LiLiu}. Few pieces of evidence exist concerning the
validity of the bound for self-gravitating systems
\cite{Zas1,Sorkin,Zurek}. However, the universal bound Eq. (\ref{Eq2})
is known to be true independently of black-hole physics for a variety
of systems in which gravity is negligible 
\cite{Beken8,BekenSchi,SchiBeken,BekenGuen}. In particular, Schiffer
and Bekenstein \cite{SchiBeken} had provided an analytic proof of the bound for free
scalar, electromagnetic and massless spinor fields enclosed in boxes
of arbitrary shape and topology.

In this paper we test the validity of the GSL 
in an idealized gedanken experiment in which an entropy-bearing {\it rotating} 
system falls into a stationary black hole. We argue that while
the bound Eq. (\ref{Eq2}) may be a necessary condition for the 
fulfillment of the GSL, it may {\it not} be a sufficient one.

It is not difficult to see why a {\it stronger} upper bound must exist
for the entropy $S$ of an arbitrary system with 
energy $E$, intrinsic angular 
momentum $s$ and (maximal) radius $R$: The {\it gravitational 
spin-orbit interaction} \cite{Wald} (the analog of the more familiar 
electromagnetic spin-orbit interaction) experienced by the 
spinning body (which, of course, was not
relevant in the above mentioned gedanken experiment) can decrease the 
energy delivered to the black hole. This would {\it decrease} 
the change in black-hole entropy (area). Hence, the GSL will be
violated unless the spinning-system entropy (what disappears from the
black-hole exterior) is restricted by a bound {\it stronger} 
than Eq. (\ref{Eq2}).

Furthermore, there is one disturbing feature of the universal 
bound Eq. (\ref{Eq2}). As was pointed out by Bekenstein \cite{Beken5},
Kerr black holes conform to the bound; however, only the
Schwarzschild hole actually {\it attains} the bound. This uniqueness of
the Schwarzschild black hole (in the sense that it is the {\it only} 
black hole which have the maximum entropy allowed by quantum theory and
general relativity) among the electrically neutral 
Kerr-family solutions is somewhat 
disturbing. Clearly, the unity of physics demands a stronger bound for
{\it rotating} systems in general, and for black holes in particular 
(see also \cite{HBML}).

In fact, the plausible existence of an upper bound stronger than
Eq. (\ref{Eq2}) on the entropy of a rotating system 
has nothing to do with black-hole physics. Classically, entropy 
is a measure of the phase space available to the system in question. 
Consider a system whose energy is no more than $E$. The limitation
imposed on $E$ amounts to a limitation on the momentum space available
to the system's components (provided the potential energy is bounded
from below). Now, if part of the system's energy is in the form of
a coherent (global) kinetic energy (in contrast to random motion of its
constituents), then the momentum space available
to the system's components is {\it further} limited (part of the
energy of the system is irrelevant for the system's statistical
properties). If the system has a
finite dimension in space, then its phase space is limited. This
amounts to an upper bound on its entropy. This bound 
evidently {\it decreases} with the absolute value of the intrinsic 
angular momentum of the system. 
However, our simple argument cannot yield the exact dependence of the
entropy bound on the system's parameters: its energy, 
intrinsic angular momentum (spin), and proper radius. 

In fact, black-hole physics (more precisely, the GSL) provides a 
concrete universal upper bound for rotating systems. 
We consider a spinning body of rest mass $\mu$, (intrinsic) spin $s$
and proper cylindrical radius $R$, which is descending into a black
hole. We consider plane (equatorial) motions of the body in a Kerr-Newman 
background \cite{MTW}, with the (intrinsic) spin orthogonal to the
plane (the general motion of a spinning particle in a 
Kerr-Newman background is very complicated, and 
has not been analyzed so far). 
The black-hole (event and inner) horizons are located at

\begin{equation}\label{Eq3}
r \pm =M \pm (M^2-Q^2-a^2)^{1/2}\  ,
\end{equation}
where $M$, $Q$ and $a$ are the mass, charge and angular-momentum per
unit mass of the hole, respectively (we use gravitational units in 
which $G=c=1$). The test particle approximation implies $|s| / (\mu r_+) \ll 1$.

The equation of motion of a spinning body in the equatorial plane
of a Kerr-Newman background is a quadratic 
equation for the conserved energy (energy-at-infinity) $E$ of 
the body \cite{Hojman}

\begin{equation}\label{Eq4}
\tilde \alpha E^2-2 \tilde \beta E + \tilde \gamma=0\  ,
\end{equation}
where the expression for $\tilde \alpha, \tilde \beta$ 
and $\tilde \gamma$ are given in \cite{Hojman}.

The actual role of buoyancy forces in the context of the GSL is 
controversial (see e.g., \cite{Beken6,Beken7,UnWal,PelWal}). 
Bekenstein \cite{Beken7} has recently shown that buoyancy
protects the GSL, provided the floating point (see
\cite{UnWal,Beken6,Beken7}) is close to the black-hole
horizon. In addition, Bekenstein \cite{Beken7} has proved that one
can derive the universal entropy bound Eq. (\ref{Eq2}) from the GSL
when the floating point is near the horizon (this is the relevant
physical situation for macroscopic and mesoscopic objects with a moderate number of
species in the radiation, which seems to be the case in our world). 
The entropy bound Eq. (\ref{Eq2}) is also a {\it sufficient} condition for
the validity of the GSL. For
simplicity, and in the spirit of the original analysis of Bekenstein
\cite{Beken5}, we neglect buoyancy contribution to the energy bookkeeping of
the body. As in the case of non rotating systems \cite{Beken7} we expect this not to
effect the final entropy bound.

The gradual approach to the black hole must stop when the
proper distance from the body's center of mass to the black-hole
horizon equals $R$, the body's radius. Thus, in order to find the
change in black-hole surface area caused by
an assimilation of the spinning body, one should first solve
Eq. (\ref{Eq4}) for $E$ and then evaluate it at the point of capture 
$r=r_{+}+ \delta (R)$, where $\delta(R)$ is 
determined by 

\begin{equation}\label{Eq5}
\int_{r_{+}}^{r_{+}+ \delta (R)} (g_{rr})^{1/2} dr = R\  ,
\end{equation}
with $g_{rr}=(r^2+a^2 cos^2 \theta) \Delta^{-1}$, 
and $\Delta =(r-r_-)(r-r_+)$. Integrating Eq. (\ref{Eq5}) one finds (for $\theta = \pi / 2$ 
and $R \ll r_{+}$)

\begin{equation}\label{Eq6}
\delta (R)=(r_{+}-r_{-}) {R^2 \over {4{r^2_+}}}\  .
\end{equation}
Thus, the conserved energy $E$ of a body having a radial turning 
point at $r=r_{+}+ \delta (R)$ \cite{note1} is 

\begin{equation}\label{Eq7}
E = {{aJ} \over \alpha} - {{Js(r_+ - r_-)r_+} 
\over {2 \mu \alpha^2}} + {{R (r_+-r_-)} \over {2 \alpha}} \sqrt{\mu^2
+J^2 {r_+^2 \over \alpha^2}}\  ,
\end{equation}
where the ``rationalized area'' $\alpha$ is 
related to the black hole surface area $A$ by
$\alpha = A/4 \pi$, and $J$ is the body's total angular momentum.
The second term on the r.h.s. of Eq. (\ref{Eq7}) represents the 
above mentioned gravitational {\it spin-orbit} interaction between the 
orbital angular momentum of the body and its
intrinsic angular momentum (spin).

An assimilation of the spinning body by the black hole results in a 
change $dM=E$ in the black-hole 
mass and a change $dL=J$ in its angular momentum. 
Using the first-law of black hole thermodynamics \cite{BarCarHaw}

\begin{equation}\label{Eq8}
dM={\kappa \over {8\pi}} dA + \Omega dL\  ,
\end{equation}
where $\kappa=(r_{+}-r_{-})/2\alpha$ and $\Omega=a/ \alpha$ are the
surface gravity ($2\pi$ times the Hawking temperature \cite{Haw3}) and
rotational angular frequency of the black hole, respectively, we find

\begin{equation}\label{Eq9}
d\alpha =-{{2Jsr_+} \over {\mu \alpha}} +2R\sqrt{\mu^2
+J^2 {r_+^2 \over \alpha^2}}\  .
\end{equation}

The increase in black-hole surface area Eq. (\ref{Eq9}) can 
be {\it minimized} if the total angular 
momentum of the body is given by

\begin{equation}\label{Eq10}
J=J^* \equiv {{s \alpha} \over {{R r_+} 
\sqrt {1 -\left({s \over {\mu R}} \right)^2}}}\  .
\end{equation}
For this value of $J$ the area increase is 

\begin{equation}\label{Eq11}
(\Delta A)_{min}=8 \pi \mu R \sqrt {1 -\left ({s 
\over {\mu R}} \right)^2}\  ,
\end{equation}
which is the minimal increase in black-hole surface area 
caused by an assimilation of a spinning body with given 
parameters $\mu, s$ and $R$. Obviously, a 
minimum exists only for $s \leq \mu R$. Otherwise, $\Delta A$ can be
made (arbitrarily) negative, violating the GSL. 
Moller's well-known theorem \cite{Moller} therefore protects the GSL.

Arguing from the GSL, we derive an {\it upper bound} to the entropy $S$ of an
arbitrary system of proper energy $E$, intrinsic 
angular momentum $s$ and proper radius $R$:

\begin{equation}\label{Eq12}
S \leq 2 \pi \sqrt {(RE)^2 -s^2}/\hbar \  .
\end{equation}
It is evident from this suggestive argument that in order for the GSL to
be satisfied $[(\Delta S)_{tot} \equiv (\Delta S)_{bh} -S \geq 0]$,
the entropy $S$ of the rotating system should be bounded as 
in Eq. (\ref{Eq12}). This upper bound is {\it universal} in the sense
that it depends only on the {\it system's} 
parameters (it is {\it independent} of the black-hole parameters which
were used to suggest it).

It is in order to emphasize an important assumption made in obtaining the
upper bound Eq. (\ref{Eq12}); We have not taken into account {\it second}-order
interactions between the particle's angular momentum and the black hole, which are
expected to be of order $O(J^2/M^3)$. Taking cognizance of
Eq. (\ref{Eq10}) we learn that this approximation is justified for
rotating systems with negligible self gravity, i.e., rotating systems
with $\mu \ll R$.

Although our derivation of the entropy bound is valid only for rotating
systems with negligible self-gravity, we {\it conjecture} that it might be
applicable also for strongly gravitating systems; A positive evidence
for the validity of the bound is the fact that any Kerr black hole
saturates it, provided the effective radius 
$R$ is properly defined for the black hole: consider an electrically 
neutral Kerr black hole. Let its energy and angular momentum be $E=M$
and $s=Ma$, respectively. The black-hole entropy
$S_{BH}=A/4\hbar=\pi(r^2_++a^2)/\hbar$ exactly saturates the entropy 
bound provided one identifies the {\it effective} radius $R$ with $(r^2_++a^2)^{1/2}$, where
$r_+=M+(M^2-a^2)^{1/2}$ is the radial Boyer-Lindquist coordinate for
the Kerr black-hole horizon. The identification may be reasonable because
$4\pi (r^2_++a^2)$ is exactly the black-hole surface area. 

Evidently, systems with negligible self-gravity (the rotating system in our
gedanken experiment) and systems with maximal gravitational effects 
(i.e., rotating black holes) both satisfy the upper bound
Eq. (\ref{Eq12}). Thus, this bound appears to be of universal
validity. The intriguing feature of our derivation is that it uses a law
whose very meaning stems from gravitation (the GSL, or equivalently
the area-entropy relation for black holes) to derive a universal 
bound which has {\it nothing} to do with 
gravitation [written out fully, the bound
Eq. (\ref{Eq12}) would involve $\hbar$ and $c$, but {\it not}
$G$]. This provides a striking illustration of the {\it unity} of physics.

In summary, an application the generalized second law of
thermodynamics to an idealized gedanken experiment in which 
an entropy-bearing rotating system falls into a black hole, enables us
to conjecture an {\it improved upper bound} to the entropy of 
a {\it rotating} system. 
The bound is stronger than Bekenstein's bound for non-rotating
systems. Moreover, this bound seems to be remarkable
from a black-hole physics point of view: provided the effective
radius $R$ is properly defined, {\it all} Kerr black
holes {\it saturate} it (although we emphasize again that our specific
derivation of the bound is consistent only for systems 
with {\it negligible} self gravity). This suggests that the Schwarzschild black
hole is {\it not} unique from a black-hole entropy point of
view, removing the disturbing feature of the entropy bound
Eq. (\ref{Eq2}). Thus, {\it all} electrically neutral black holes 
seem to have the {\it maximum} entropy allowed by quantum theory 
and general relativity. This provides a striking illustration of 
the {\it extreme} character displayed by (all) black holes, which is,
however, still {\it within} the boundaries of more mundane physics.

\bigskip
\noindent
{\bf ACKNOWLEDGMENTS}
\bigskip

I thank Jacob D. Bekenstein for helpful discussions. 
This research was supported by a grant from the Israel Science Foundation.


\begin{thebibliography}{99}
\bibitem{note1} As our task is to test the validity of the GSL in 
the most 'dangerous' situation [with $(\Delta S)_{bh}$ as small as 
possible] we consider the case of a spinning body which is 
captured from a radial turning point of its motion. This minimize the 
increase in black hole surface area, and thus, allows one to derive the 
strongest bounds on the various physical quantities.
\end{thebibliography}
\end{document}